\documentclass[a4paper,12pt]{article}
\usepackage {amsmath}
\usepackage {amsxtra}
\usepackage {amsbsy}
\usepackage {amscd}
\usepackage {latexsym}
\usepackage {amssymb}
\usepackage {amsfonts}
\usepackage {a4wide}
\usepackage {indentfirst}
\usepackage {cite}

\newcommand {\oks}[2]{{\raise0.7ex\hbox{${\scriptstyle #1}$}\!\mathord{\left/
{\vphantom{{1}{2}}}\right.\kern-\nulldelimiterspace}\!\lower0.7ex
\hbox{${\scriptstyle #2}$}}}

\newcommand {\bb}[1]{\mbox{\boldmath $#1$}}

\begin{document}

\title{Neutrino self-polarization effect in matter}
\author{Andrey Lobanov,
  Alexander Studenikin\thanks{E-mail:
  studenik@srd.sinp.msu.ru}
  \\
  Moscow State University,
  \\
  Department of Theoretical physics,
  \\
  119992  Moscow, Russia}

  \date{}
  \maketitle

\begin{abstract}{The quasi-classical theory of the
spin light of neutrino ($SL\nu$) in background matter, accounting
for the neutrino polarization, is developed. The neutrino
transitions $\nu_{L}\rightarrow \nu_{R}$ and $\nu_{R}\rightarrow
\nu_{L}$ rates in matter are calculated. It is shown that  the
$SL\nu$ in matter leads to the neutrino conversion from active
$\nu_{L}$ to sterile $\nu_{R}$ states (neutrino self-polarization
effect in matter)}.
\end{abstract}

Convincing evidence in favour of non-zero neutrino mass that has
been obtained during the last few years in atmospheric and
solar-neutrino experiments, are also confirming in the reactor
KamLAND and long-baseline accelerator experiments (see
\cite{Bil04}  for a review on the present status of neutrino
mixing and oscillations). Even within the standard model
(minimally extended with $SU(2)-$singlet right-handed neutrino) a
massive neutrino inevitably has non-zero magnetic moment $\mu$
generated by the one-loop diagramme \cite {FujShr80}. A recent
studies of a massive neutrino electromagnetic properties within
one-loop level, including discussion on the neutrino magnetic
moment, can be found in \cite{DvoStuPRD04}. It should be also
noted here that a rather detailed discussion on the neutrino
charge radius is presented in the two recent papers
\cite{BerPapBin0405288,FujShrPRD04}.

In a series of our papers \cite{LikStu95, EgoLobStuPLB00,
LobStuPLB01, DvoStuJHEP02,
%GriLobStuPLB02,
LobStuPLB03,DvoGriStu04
%StuYF04
} we have developed the Lorentz invariant approach to neutrino
oscillations which enables us to study, in particular, the
neutrino spin procession in the background matter with effects of
the presence of electromagnetic and gravitational fields being
also accounted for. A review on these our studies can be found in
\cite{StuYF04Stu0407010}.

In \cite{LobStuPLB03,DvoGriStu04}  we have predicted the new
mechanism of electromagnetic radiation by neutrino moving in
background matter and/or electromagnetic and gravitational fields.
We have named this radiation  as ``spin light of neutrino'' and
introduced the abbreviation $SL\nu$ which we shall use below in
this paper. The $SL\nu$ originates from the neutrino spin
precession that can be induced whether by weak interactions of
neutrino with the background matter or by the external
electromagnetic or gravitational fields that could be present in
the background environment. It should be noted  that the discussed
mechanism of electromagnetic radiation by a neutrino moving in a
constant magnetic field was also studied previously in
\cite{BorZhuTer88}.

As we have shown in \cite{LobStuPLB03}, the total power of the
$SL\nu$ in matter does not washed out when the emitted photon
refractive index is equal to unit and the $SL\nu$ can not be
considered as the neutrino Cerenkov radiation (see, for example,
\cite{IoaRaf97} and references therein). It was also emphasized
\cite{LobStuPLB03} that the initially unpolarized neutrino beam
(equal mixture of active left-handed and sterile right-handed
neutrinos) can be converted to the totally polarized beam composed
of only $\nu_{R}$ due to the spin light in contrast to the
Cherenkov radiation which can not produce the neutrino spin
self-polarization effect.

The discovered important properties of $SL\nu$ (such as strong
beaming of the radiation along the neutrino momentum, the rapid
growth of the total radiation power with the neutrino energy and
density of matter, the possibility to emit photons with energies
span up to gamma-rays) enables us to predict that this radiation
should be important in different astrophysical environments
(quasars, gamma-ray bursts etc) and in dense plasma of the early
Universe.

In this paper we should like to present a detailed study of the
neutrino spin self-polarization effect in matter which has been
recently predicted in our previous paper \cite{LobStuPLB03}. In
\cite{LobStuPLB03} we considered the $SL\nu$ in matter in the case
of {\it unpolarized}  neutrinos. In the present paper we make a
reasonable step forward and  study the $SL\nu$ in matter
accounting for the neutrinos polarization. It is sufficient that
the $SL\nu$ in matter (as the similar radiation by neutrinos
moving in the magnetic field \cite{BorZhuTer88}) originates from
the spin-flip transitions $\nu_{L} \rightarrow \nu_{R}$. Within
the quantum approach the corresponding Feynman diagram of the
proposed new process is the standard one-photon emission diagram
with the initial and final neutrino states described by the "broad
lines" that account for the neutrino interaction with matter.  We
show below how derive the transition rate of the {\it polarized}
neutrinos using the quasi-classical method \cite{Lob03} for
description of spin wave functions in the presence of an
electromagnetic field, given by the tensor $F_{\mu\nu}$, and apply
it for the case when a massive neutrino with non-zero magnetic
moment is moving and radiating the spin light in the background
matter.

As it has been shown \cite{EgoLobStuPLB00}, the quasi-classical
Bargmann-Michel-Telegdi equation \cite{BarMicTelPRL59}, describing
a neutral particle spin evolution under the influence of an
electromagnetic field, can be generalized for the case of a
neutrino moving in electromagnetic fields and matter by
implementing the substitution  of the external electromagnetic
field tensor, $F_{\mu \nu}=({\bf E}, {\bf B})$, according to the
prescription
\begin{equation}\label{presc}
F_{\mu\nu} \rightarrow F_{\mu\nu}+G_{\mu\nu},
\end{equation}
where the anti-symmetric tensor $G_{\mu \nu}=(- {\bf P}, {\bf M})$
describes the neutrino interaction with the background matter. We
have also shown \cite{LobStuPLB01} how to construct the tensor
$G_{\mu \nu}$ with use of the neutrino speed, matter speed, and
matter polarization four-vectors. It worth to note that the
substitution (\ref{presc}) implies that the magnetic ${\bf B}$ and
electric ${\bf E}$ fields are shifted by the vectors ${\bf M}$ and
${\bf P}$, respectively:
\begin{equation}
{\bf B} \rightarrow {\bf B} +{\bf M}, \ \ {\bf E} \rightarrow {\bf
E} -{\bf P}. \label{11}
\end{equation}
From the new generalized BMT equation for the neutrino spin
evolution in an electromagnetic field and matter we have finally
derive  the following equation for the evolution of the
three-di\-men\-sio\-nal neutrino spin vector ${\bf S}$ (see
\cite{LobStuPLB01,LobStuPLB03}):

\begin{equation}\label{S} {d{\bf S}
\over dt}={2\mu \over \gamma} \Big[ {{\bf S} \times ({{\bf
B_0}}+{\bf M_0})} \Big],
\end{equation}
\begin{equation}
{\bf B_0} =\gamma\Big({\bf B_{\perp}} +{1 \over \gamma}{\bf
B_{\parallel}} +  \Big[{{\bf E_{\perp}} \times {\bb
\beta}}\Big]\Big), \ \gamma = (1-\beta^2)^{-{1 \over 2}},
\end{equation}

\begin{equation}
{\bf {M_0}}={\bf {M}{_{0_{\parallel}}}}+ {\bf {M_{0_{\perp}}}},
\label{M_0}
\end{equation}
\begin{equation}
\begin{array}{c}
\displaystyle {\bf {M}_{0_{\parallel}}}=\gamma{\bb\beta}{n_{0}
\over \sqrt {1- v_{e}^{2}}}\left\{ \rho^{(1)}_{e}\left(1-{{{\bf
v}}_e {\bb\beta} \over {1- {\gamma^{-2}}}} \right)\right. \\-
\displaystyle\rho^{(2)}_{e}\left. \left({\bb \zeta_{e}}{\bb\beta}
\sqrt{1-v^2_e}+ {({\bb \zeta}_{e}{{\bf v}}_e)({\bb\beta}{{\bf
v}}_e) \over 1+\sqrt{1-v^2_e} }\right){1 \over {1- {\gamma^{-2}}}}
\right\}, \label{M_0_parallel}
\end{array}
\end{equation}
\begin{equation}\label{M_0_perp}
\begin{array}{c}
\displaystyle {\bf {M}}_{0_{\perp}}=-\frac{n_{0}}{\sqrt {1-
v_{e}^{2}}}\Bigg\{ {\bf{v}}_{e_{\perp}}\Big(
\rho^{(1)}_{e}+\displaystyle\rho^{(2)}_{e}\frac
{({\bb{\zeta}}_{e}{{\bf{v}}_e})} {1+\sqrt{1-v^2_e}}\Big)+
{{\bb{\zeta}}_{e_{\perp}}}\rho^{(2)}_{e}\sqrt{1-v^2_e}\Bigg\},
\end{array}
\end{equation}
where $t$ is time in the laboratory frame, ${\bb \beta}$ is the
neutrino speed, ${\bf F}_{\perp}$ and ${\bf F}_{\parallel}$ (${\bf
F}= {\bf B},{\bf E}$)  are transversal and longitudinal (with
respect to the direction ${\bf n}$ of neutrino motion)
electromagnetic field components in the laboratory frame. For
simplify  we neglect here the neutrino electric dipole moment,
$\epsilon=0$, and also consider the case when matter is composed
of only one type of fermions (which we chose to be electrons for
definiteness). Here $n_0=n_{e}\sqrt {1-v^{2}_{e}}$ is the
invariant number density of matter given in the reference frame
for which the total speed of matter is zero. The vectors ${\bf
v}_e$, and ${\bb \zeta}_e \ (0\leqslant |{\bb \zeta}_e |^2
\leqslant 1)$ denote, respectively, the speed of the reference
frame in which the mean momentum of matter (electrons) is zero,
and the mean value of the polarization vector of the background
electrons in the above mentioned reference frame. The coefficients
$\rho^{(1,2)}_e$ are calculated if the neutrino Lagrangian is
given, and  within the extended standard model supplied with
$SU(2)$-singlet right-handed neutrino $\nu_{R}$,
\begin{equation}\label{rho}
\rho^{(1)}_e={\tilde{G}_F \over {2\sqrt{2}\mu }}\,, \qquad
\rho^{(2)}_e =-{G_F \over {2\sqrt{2}\mu}}\,,
\end{equation}
where $\tilde{G}_{F}={G}_{F}(1+4\sin^2 \theta _W).$

Our observation \cite{EgoLobStuPLB00,LobStuPLB01} that the
neutrino spin evolution in the presence of matter can be described
by the generalized BMT equation with the substitutions given by
eqs.(\ref{presc}) and (\ref{11}) make it possible to use the
quasi-classical approach (developed in \cite{Lob03}  for the case
of a neutral particle spin evolution in electromagnetic fields)
for the study of the neutrino spin-polarization effect in matter.
Here below we suppose that the effect of an external
electromagnetic field (if any is present in the background
environment) can be neglected in comparison with the effect of the
neutrino interaction with the background matter. Then, the
equation for the neutrino quasi-classical spin wave function
$\Psi(\tau)$ is

\begin{equation}\label{spin_eq}
i{d\Psi \over d\tau}={1 \over 2}\mu \epsilon_{\mu \nu \rho
\lambda}
 G^{\rho \lambda}u^{\nu} \gamma^{\mu}\gamma^{5} \hat{u}\Psi,
\end{equation}
where $\tau$ is the neutrino proper time and we use the notation
$\hat{u}=\gamma_{\mu}u^{\mu}$, where $u^{\mu}=(\gamma, \gamma
{\bb{\beta}}).$ If for simplicity we neglect effects of the
neutrino electric dipole moment ($\epsilon=0$) and consider
unpolarized matter composed of electrons, then we
have\cite{LobStuPLB03}
\begin{equation}\label{G}
G^{\mu \nu}= \epsilon^{\mu \nu \rho \lambda} j_{\rho}
u_{\lambda}\rho^{(1)},
\end{equation}
where
\begin{equation}\label{j}
j^\mu=(n,n{\bf {v}}),
\end{equation}
is the electron current, $\bf {v}$ is the speed of matter (the
average speed of the electrons), and we use the notation
$n=\frac{n_{0}}{\sqrt{1-v_{e}^{2}}}$. In the case of non-moving
matter that will be also considered below, for the tensor $G^{\mu
\nu}$ we get
\begin{equation}\label{G_1}
G^{\mu \nu}= \gamma \rho^{(1)} n
\begin{pmatrix}{0}&{0}& {0}&{0} \\
{0}& {0}& {-\beta_{3}}&{\beta_{2}} \\
{0}&{\beta_{3}}& {0}&{-\beta_{1}} \\
{0}&{-\beta_{2}}& {\beta_{1}}& {0}
\end{pmatrix}.
\end{equation}

To derive the total transition probability of the neutrino spin
light radiation in matter we define the density matrix of
partially polarized neutrino in the form,
\begin{equation}\label{rho}
 \varrho(\tau,\tau') =\frac{1}{2}\,U(\tau,\tau_{0}) \big(\hat
 {p}(\tau_{0})+m_{\nu}\big)
 \big(1-\gamma^5\hat
{S}(\tau_{0})\big)U^{-1}(\tau',\tau_{0}),
\end{equation}

\noindent where $p(\tau_{0})$ is the neutrino initial momentum,
$U(\tau,\tau_{0})$ is the neutrino spin evolution operator
correspondent to the equation (\ref{spin_eq}). For the pure state
the density matrix reduces to direct product of bispinors,
normalized by the condition $\bar{\varPsi}(\tau)\varPsi(\tau) =
2m_{\nu}.$

The total transition probability of the neutrino spin light
radiation in matter is
{\setlength{\multlinegap}{2pt}\begin{multline}\label{ar1}
  \displaystyle P=-\!\!\int\!\! d^{4}x\,d^{4}y\!\int
  \frac{d^{4}p\,d^{4}q\,d^{4}k}{(2\pi)^{6}}\,
\delta(k^{2})\,\delta(p^{2}-m^{2}_{\nu})\,
 \delta(q^{2}-m^{2}_{\nu})\\\!\!\!\times\varrho^{\mu\nu}_{ph}(x,y;k)\:
{\mathrm{S p}}\big\{\varGamma_{\mu}(x)\varrho_{i}(x,y;p)
\varGamma_{\nu}(y)\varrho_{f}(y,x;q)\big\}.
\end{multline}}

\noindent Here $\varrho_{i}(x,y;p),\;\varrho_{f}(y,x;q)$ are
density matrices of the initial, $i$, and final, $f$, neutrino
states, $\,\varrho^{\mu\nu}_{ph}(x,y;k)$ is the density matrix of
the emitted photon, $\varGamma^{\mu} = -\,\sqrt{4\pi} \mu
\sigma^{\mu\nu}k_{\nu}$ is the vertex function,
$\sigma^{\mu\nu}=\frac{1}{2}(\gamma^{\mu}\gamma^{\nu}-
\gamma^{\nu}\gamma^{\mu})$.

In order to transit to the quasi-classical approximation, it is
necessary to substitute precise density matrices for once of
(\ref{rho}) and to neglect the recoil in the photon radiation
process.

After calculations similar to that performed in \cite{Lob03} for
the case of a neutral fermion transition under the influence of an
electromagnetic field, we get the quasi-classical expression for
the neutrino total transition probability in matter,
\begin{equation}\label{P_1}
P=\frac{\mu^{2}}{4\pi^{2}} \int\!\! d\Omega\!\!
\int\limits_{0}^{\infty}\!k^{3}dk\!\iint\!d\tau
d\tau'e^{ik(lu)(\tau-\tau')}\,T(\tau,\tau';u),
\end{equation}
%\begin{equation}\label{prob1}
%  P={\mu \over {(2\pi)^{2}}} \int d\Omega \int_{0}^{\infty} k^3 dk
%  \int d\tau d\tau' e^{ik(lu)(\tau - \tau')}T(\tau,\tau';u),
%\end{equation}
where
\begin{equation}\label{T}
T(\tau,\tau';u)=V_{i}V_{f} - A_{i}A_{f},
\end{equation}
and
\begin{equation}\label{ar8}
\begin{array}{l}
\displaystyle  V_{i,f}\!=\! \frac{1}{4}{\mathrm {Sp}}\!\left\{
\hat{l}\,U(\tau)(1+\hat{u})
  (1\!-\gamma^{5}{\hat{S}}_{0i,f})U^{-1}(\tau')\right\}\!,\\[12pt]
\displaystyle A_{i,f}\!= \!\frac{1}{4}{\mathrm
{Sp}}\Big\{\!\gamma^{5} \hat{l}\,U(\tau)(1+\hat{u})
  (1\!-\gamma^{5}{\hat{S}}_{0i,f})U^{-1}(\tau')\Big\}.
\end{array}
\end{equation}
Integrations in eq.(\ref{P_1}) are performed over the solid angle
$\Omega$ and energy $k$ of the photon, and the proper times $\tau$
and $\tau'$ of neutrino. The four-dimensional vector $l^\nu=\{
1,{\bf l}\}$ is fixed by the three-dimensional unit vector ${\bf
l}$ that points the direction of radiation.

Performing integration over angular variables, we get from
(\ref{P_1}) the neutrino transition probability in the form
\begin{equation}\label{P_2}
  P=\frac{4\mu^{2}}{3\pi}\!\!
\iint\!d\tau d\tau'\frac{1}{2(\tau\!-\tau'\!+i0)}
  (\partial_{\tau}\partial_{\tau'}^{2}-
  \partial_{\tau}^{2}\partial_{\tau'})\tilde{V}_{i}\tilde{V}_{f},
\end{equation}
\noindent where
\begin{equation}\label{ar013}
\tilde{V}_{i,f}= \frac{1}{4}{\mathrm {Sp}}\left\{ U(\tau)
  (1-\gamma^{5}{\hat{S}}_{0i,f})U^{-1}(\tau')\right\}.
\end{equation}

Let us consider a neutrino propagating in matter composed of
unpolarized electrons. In this case the tensor $G_{\mu \nu}$ is
given by (\ref{G}). Then the neutrino spin evolution operator
$U(\tau)$ correspondent to the equation (\ref{spin_eq}) is
\begin{equation}\label{ar0240}
\displaystyle  U(\tau,\tau_{0})=\cos {\omega}(\tau -\tau_{0}) +
i\gamma^{5} \hat{S}_{tp} \hat{u}\sin
  {\omega}(\tau -\tau_{0}),
\end{equation}
where
\begin{equation}\label{S_tp}
S^{\mu}_{tp}= -\,\frac{{\textsl{j}^\mu} -u^{\mu}(u\textsl{j})}
{\sqrt{(uj)^{2}-(\textsl{j})^{2}}}
\end{equation}
is the four-dimensional vector determining the axis of the total
neutrino  spin self-polarization. The neutrino spin precession
frequency $\omega$ is determined by the neutrino speed vector
$u_{\alpha}$ and the tensor $G_{\mu \nu}$:
\begin{equation}\label{omega}
\omega=\displaystyle\mu\sqrt{u_{\alpha}G^{\alpha \mu}G_{\mu
\nu}u^{\nu}}.
\end{equation}
In the case of moving and unpolarized matter composed of electrons
we have from eqs.(\ref{G}) and (\ref{j}) that
\begin{equation}\label{omega_1}
\omega=\mu\rho^{(1)}n\gamma \sqrt{(1-{\bf v} {\bb \beta})^{2} -
(1-v^{2})/{\gamma}}.
\end{equation}
Note that the latter expression for the frequency $\omega$ in the
case of non-moving matter is in agreement with the estimation of
ref.\cite{LobStuPLB03} for the neutrino spin light photon energy
\begin{equation}
\omega_{\gamma}\sim G_{F}n\gamma^{2}
\end{equation}
 in the laboratory reference frame.

From the previous discussion it is evident that the neutrino spin
polarization axis in the rest frame of the neutrino is given by
the vector ${\bf M_0}$ (see eqs.(\ref{M_0}),(\ref{M_0_parallel})
and (\ref{M_0_perp})). It follows from (\ref{S_tp}) that in the
case of unpolarized and non-moving matter the direction of the
neutrino spin polarization coincides with the direction of the
neutrino speed,
\begin{equation}
{\bf S}_{tp}={\bb \beta} / \beta.
\end{equation}
The neutrino spin light radiation leads to the total spin
polarization in the direction of the neutrino motion, i.e.
initially left-handed polarized neutrinos are converted to the
right-handed polarized neutrinos,
\begin{equation}
\nu_{L} \rightarrow \nu_{R}.
\end{equation}
From (\ref{P_2}) we get that the neutrino transition rate (the
probability per unite time) from $\nu_{L}$ to $\nu_{R}$ state is
\begin{equation}\label{Gamma_L_R}
  \Gamma_ {\nu_{L}\rightarrow \nu_{R}}= \frac{32}{3
}\mu^{2}\gamma^{-1}\omega^{3},
\end{equation}
\noindent whereas the rate of the transition
$\nu_{R}\rightarrow\nu_{L}$ is zero,
\begin{equation}\label{Gamma_R_L}
  \Gamma_ {\nu_{R}\rightarrow \nu_{L}}= 0.
\end{equation}
It should be noted here that eqs. (\ref{Gamma_L_R}) and
(\ref{Gamma_R_L}) with $\omega$ determined by (\ref{omega})
reproduce the neutrino transition rates for the unpolarized
background matter of arbitrary particles composition moving with
arbitrary common speed if the appropriate form (see in
\cite{LobStuPLB01}) of the tensor $G_{\mu \nu}$ is chosen. If
matter is not moving and composed of only electrons then from eqs.
(\ref{rho}), (\ref{omega}), (\ref{omega_1}) and (\ref{Gamma_L_R})
we get
\begin{equation}\label{Gamma_L_R_1}
 \Gamma_ {\nu_{L}\rightarrow \nu_{R}}=
\frac{2\sqrt{2}}{3 }\mu^{2}\gamma^{2}{\tilde{G}_{F}}^{3}n^{3},
\end{equation}
and, obviously, the rate of the transition
$\nu_{R}\rightarrow\nu_{L}$ is again zero.  The obtained result
(\ref{Gamma_L_R_1}) exceeds the value of the neutrino spin light
rate derived in \cite{LobStuPLB03} by a factor of two because it
gives the emission rate of the totaly polarized left-handed
neutrinos $\nu_{L}$, whereas the corresponding rate of
ref.\cite{LobStuPLB03} was derived under the assumption that
neutrinos in the initial state were not polarized.

For the ultra relativistic neutrino (that is the most interesting
case for different astro\-physical and cosmology applications)
interactions of the right-handed polarized neutrinos, $\nu_{R}$,
with the background particles is suppressed with respect to
interactions of the left-handed polarized neutrinos, $\nu_{L}$, by
the factor of $\sim\gamma ^{-1}$. Therefore, we conclude that, in
fact, the ``spin light of neutrino'' in matter leads to the
neutrino conversion from active to sterile states. As it follows
from the above discussion, the rate of the ``spin light of
neutrino'' significantly depends on the density of the background
matter. That is why the considered neutrino self-polarization
effect is expected to be important at the early stages of
evolution of the Universe.

\end{document}